\documentstyle[prl,aps,twocolumn]{revtex}

\begin{document}

\title{
Actions and 
symmetries of NSR superstrings and D-strings
}

\author{
Friedemann Brandt\,$^{a,b}$, Alexander Kling\,$^c$, Maximilian Kreuzer\,$^c$
}

\address{ 
$^a$\,Max-Planck-Institute for Mathematics in the Sciences,
Inselstr.\ 22-26, D-04103 Leipzig, Germany\\
$^b$\,Institut f\"ur Theoretische Physik, Universit\"at Hannover,
Appelstr.\ 2,
D-30167 Hannover, Germany\\
$^c$\,Institut f\"ur Theoretische Physik, Technische Universit\"at Wien,\\
Wiedner Hauptstra\ss e 8-10, A-1040 Vienna, Austria
\\[1.5ex]
\begin{minipage}{14cm}\rm\quad
We present all NSR superstring and super-D-string actions invariant under 
a set of prescribed gauge transformations, and characterize completely 
their global symmetries. In particular we obtain locally supersymmetric 
Born-Infeld actions on general backgrounds in a formulation with extra
target space dimensions. The nontrivial global symmetries of the superstring 
actions correspond to isometries of the background, whereas super-D-string 
actions can have additional symmetries acting nontrivially also on the 
coordinates of the extra dimensions.
\\[1ex]     
\end{minipage}
}

\maketitle

This letter is devoted to locally supersymmetric world-sheet actions for 
superstrings and super-D-strings and to their symmetries.
For the standard spinning string in a flat background \cite{RNS}, such an 
action has been known for more than 20 years \cite{DZBDH}. It played an 
important r\^ole in string theory, in particular as starting point for 
covariant quantization, and was extended to general backgrounds \cite{BRSSS}.
In this letter we present analogous actions for the super-D-string, i.e.\ 
locally supersymmetric world-sheet actions of the Born-Infeld type.
To our knowledge, such actions have not been explicitly constructed 
previously, except for the D2-brane \cite{BHL}. In addition we characterize 
completely the global symmetries of these actions.

As it is well known the tension of a super-p-brane may be generated 
dynamically as the flux of a world-volume p-form gauge field \cite{BLK}.
Since D-branes are described by 
Born-Infeld type actions it is natural to construct a manifestly S-dual 
IIB superstring by assembling the Born-Infeld field and a tension generating
gauge field into an $SL(2)$ doublet \cite{CT}. The resulting theory is
in a sense twelve-dimensional. This is quite
apparent in our formulation which uses actually a twelve-dimensional
target space. The two ``extra dimensions'' get frozen: 
the values of their coordinates 
are integration constants by the equations of motion.
In the light of this interpretation in terms of extra dimensions, it is
quite interesting that the symmetries of the super-D-string action are
not exhausted by the standard isometries of the ``ordinary''
(ten-dimensional) target space: rather, there can be additional symmetries
which act nontrivially also on the coordinates of the extra dimensions.
Our results may thus be of interest also in view of a possible 
twelve-dimensional theory underlying IIB superstring theory \cite{CMHCV}. 
Inspired by such considerations (and for the sake of generality) we allow 
for an arbitrary number of gauge fields and corresponding ``extra 
dimensions'', thereby including D-strings but not restricting to them.

\paragraph{Field content and gauge symmetries.}
The results reported here were obtained by a BRST-cohomological
analysis whose details will be presented elsewhere. The input of this 
analysis is the field content and a set of gauge transformations of the 
fields. The fields are those of the $2d$ supergravity multiplet $\{e_m^a,
\chi^\alpha_m,S\}$ ($m,a,\alpha$ are $2d$ world-sheet, Lorentz and spinor 
indices, respectively), an arbitrary number of ``matter multiplets'' 
$\{X^M,\psi^M_\alpha,F^M\}$ and an arbitrary number of gauge multiplets 
$\{A_m^i,\lambda^i_\alpha,\phi^i\}$. $e_m^a$ and $\chi^\alpha_m$ are the 
world-sheet zwei\-bein and gravitino, respectively, the $A_m^i$ are 
abelian gauge fields, and the $X^M$ may be regarded as coordinates
of an enlarged target space. 
All fields, including the auxiliary fields are real
(we use a Majorana-Weyl basis for the $\gamma$-matrices and the signature of 
the world-sheet metric is $\eta_{ab}=\mathrm{diag}(1,-1)$; furthermore 
$\varepsilon^{01}=\varepsilon_{10}=1$ and $\gamma_*$ is defined through
$\gamma^a\gamma^b=\eta^{ab}{1\hspace{-0.243em}\mathrm{l}} 
+ \varepsilon^{ab}\gamma_*$; our formulae apply with 
appropriate redefinitions also to Euclidean signature). 
We impose the following gauge symmetries: world-sheet diffeomorphisms, 
local (1,1) world-sheet supersymmetry, local $2d$ Lorentz invariance, 
invariance under abelian gauge transformations of the $A_m^i$,
Weyl and super-Weyl invariance, and invariance under arbitrary local shifts 
of the field $S$. 
The corresponding gauge transformations of the fields are,
written as BRST transformations,
\begin{eqnarray}
se_{m}^{~a} & = & \xi^{n}\partial_{n}e_{m}^{~a}+(\partial_{m}\xi^{n})e_{n}^{~a}
        + C_{b}^{~a}e_{m}^{~b}+C^W e_{m}^{~a} \nonumber
\\
        & &- 2{\mathrm{i}}\,\xi^{\alpha}\chi_{m}^{~\beta}
           (\gamma^a C)_{\alpha\beta} \nonumber
\\
s\chi_{m}^{~\alpha} &=& \xi^{n}\partial_{n}\chi_{m}^{~\alpha}
          +(\partial_{m}\xi^{n})\chi_{n}^{~\alpha}
          - \mbox{$\frac {1}{4}$}C^{ab}\chi_{m}^{~\beta}
          \varepsilon_{ab}(\gamma_{*})_{\beta}^{~\alpha}\nonumber
\\
         & &    + \mbox{$\frac 12$} C^W \chi_{m}^{~\alpha}
                + {\mathrm{i}} \eta^{\beta}(\gamma_{m})_{\beta}^{~\alpha} 
\nonumber 
\\
            & &+ \partial_{m}\xi^{\alpha}
         + \mbox{$\frac {1}{4}$} \xi^{\beta}(\omega_{m}^{~ab}\varepsilon_{ab}
                 (\gamma_{*})_{\beta}^{~\alpha}
                -(\gamma_m)_{\beta}^{~\alpha}S)
\nonumber 
\\
sS & = & \xi^{n}\partial_{n}S-C^W S+W 
    + {\mathrm{i}}\xi^{\gamma}(\gamma^{m}C)_{\gamma\alpha}\chi_{m}^{~\alpha}S 
\nonumber
\\
   & &  - 4\xi^{\gamma}(\gamma_{*}C)_{\gamma\alpha}\varepsilon^{nm}
        (\partial_{n}\chi_{m}^{~\alpha}
        + \mbox{$\frac {1}{4}$} 
          \omega_{n}^{~ab}\varepsilon_{ab}\chi_m^{~\beta}
           (\gamma_{*})_{\beta}^{~\alpha})
          \nonumber
\\
sX^{M} & = & \xi^{m}\partial_{m}X^{M}+\xi^{\alpha}\psi_{\alpha}^{M} \nonumber 
\\
s\psi_{\alpha}^{M} & = & \xi^{m}\partial_{m}\psi_{\alpha}^{M}
      + \mbox{$\frac {1}{4}$}C^{ab}
      \varepsilon_{ab}(\gamma_{*})_{\alpha}^{~\beta}\psi_{\beta}^{M}
           - \mbox{$\frac 12$} C^W \psi_{\alpha}^{M}
\nonumber
\\
     & &   +\xi^{\beta}C_{\beta\alpha}F^{M}  
       - {\mathrm{i}} \xi^{\beta}(\gamma^m C)_{\beta\alpha}(\partial_{m}X^{M}
       -\chi_{m}^{~\gamma}\psi_{\gamma}^{M}) 
            \nonumber 
\\
sF^{M} & = & \xi^{m}\partial_{m}F^{M}-C^W F^{M} 
           + \xi^{\alpha}(\gamma^{m})_{\alpha}^{~\beta}
           \big{\{}\partial_{m}\psi_{\beta}^{M} 
\nonumber\\
    & &   - \mbox{$\frac {1}{4}$}\omega_{m}^{~ab}
         \varepsilon_{ab}(\gamma_{*})_{\beta}^{~\gamma}\psi_{\gamma}^{M}
          - \chi_{m}^{~\gamma}C_{\gamma\beta}F^{M}
\nonumber\\
    & &     + {\mathrm{i}} \chi_{m}^{~\gamma}(\gamma^n C)_{\gamma\beta}
            (\partial_{n}X^{M}     
            - \chi_{n}^{~\delta}\psi_{\delta}^{M})
             \big{\}}\nonumber
\\
sA^i_{m} &=& \xi^{n}\partial_{n}A^i_{m}+(\partial_{m}\xi^{n})A^i_{n}
       +\partial_{m}C^i \nonumber 
\\
       & & -2{\mathrm{i}}\,\xi^{\alpha}\chi_{m}^{~\beta}
           (\gamma_* C)_{\beta\alpha}\phi^i
           - \xi^{\alpha}(\gamma_m)_{\alpha}^{~\beta}\lambda^i_{\beta} \nonumber
\\
s\phi^i &=& \xi^{n}\partial_{n}\phi^i - C^W \phi^i
           + \xi^{\alpha}(\gamma_{*})_{\alpha}^{~\beta}\lambda^i_{\beta} 
\nonumber
\\
s\lambda^i_{\beta}&=& \xi^{n}\partial_{n}\lambda^i_{\beta}
             + \mbox{$\frac {1}{4}$} C^{ab}\varepsilon_{ab}
             (\gamma_{*})_{\beta}^{~\gamma}\lambda^i_{\gamma} 
             - \mbox{$\frac {3}{2}$} C^W \lambda^i_{\beta}\nonumber
\\
       & &   + {\mathrm{i}} \xi^{\alpha}{\big \{}
             - (\gamma_{*}\gamma^{m}C)_{\alpha\beta}
               (\partial_m\phi^i-\chi_m \gamma_* \lambda^i )
\nonumber\\
       & &   + (\gamma_{*}C)_{\alpha\beta}\varepsilon^{mn}
        (\partial_m A^i_n+\chi_m\gamma_n\lambda^i-{\mathrm{i}}\chi_n
         \gamma_*C\chi_m\phi^i) 
\nonumber\\
       &  &   
             + (\gamma_{*}C)_{\alpha\beta}S\phi^i {\big \}} 
             + 2\eta_{SW}^{\alpha}(\gamma_{*}C)_{\alpha\beta}\phi^i 
\label{1}
\end{eqnarray}
where $\xi^m$ are the ghosts of world-sheet diffeomorphisms, $\xi^{\alpha}$ are
the supersymmetry ghosts, $C^{ab}$ is the Lorentz ghost, $C^W$ and 
$\eta^{\alpha}_{SW}$ are the Weyl and super-Weyl ghosts, respectively, 
$C^i$ are the ghosts associated with the gauge fields $A_m^i$, and $W$ is 
the ghost corresponding to the local shifts of the auxiliary field $S$. 
A $C$ without any index denotes the charge conjugation matrix. These gauge 
transformations were obtained from an analysis of the $2d$ supergravity 
algebra in presence of the matter and gauge multiplets \cite{K} 
(the analysis is analogous to the superspace analysis in \cite{H}).
The corresponding BRST transformations of the ghosts are such that $s^2=0$. 

\paragraph{Gauge invariant actions.}
Owing to the use of auxiliary fields, the algebra of the gauge transformations
closes off-shell. As a consequence, neither the BRST transformations 
(\ref{1}), nor the BRST transformations of the ghosts contain antifields.
This allows one to determine the action functionals which are invariant under 
the gauge transformations (\ref{1}) by computing the cohomology of $s$ in 
the space of antifield independent local functionals with ghost number 0. 
Our result of this computation is the following. The most general 
Lagrangian $L$ which transforms under the above gauge transformations into 
a total derivative and which is polynomial in the derivatives of the
fields is, modulo total derivatives,
\begin{eqnarray}
L/e&=& \mbox{$\frac {1}{2}$}
       \partial_m X^M\partial_n X^N (-h^{mn}G_{MN}+\varepsilon^{mn}B_{MN}) 
\nonumber
\\
     & &  + \mbox{$\frac {{\mathrm{i}}}{2}$} 
            \overline\psi^M\gamma^m\partial_m\psi^N G_{MN}
          + \mbox{$\frac {1}{2}$} F^M F^N G_{MN} \nonumber
\\
     & & + \chi_k\gamma^n \gamma^k (\psi^N\partial_n X^M
       -\mbox{$\frac {1}{4}$} C\chi_n \overline\psi^M\psi^N)G_{MN} \nonumber
\\  
     & &  + (\mbox{$\frac {1}{2}$} F^M \overline\psi^K \psi^N  
         -{{\mathrm{i}}}\,\overline\psi^N\gamma^m\psi^M \partial_m X^K) 
         \Gamma_{NKM} \nonumber
\\
     & &  +\mbox{$\frac {1}{4}$}(F^M \overline\psi^K \gamma_* \psi^N
          - {\mathrm{i}} \overline\psi^N\gamma^m\gamma_*\psi^M \partial_m X^K)
          H_{NKM} \nonumber
\\
     & &  - \mbox{$\frac {{\mathrm{i}}}{12}$} 
          \chi_m\gamma^n\gamma^m\psi^M\overline\psi^N\gamma_n \gamma_*\psi^K 
          H_{MNK} \nonumber
\\
     & &  +\mbox{$\frac {1}{16}$}
          \overline\psi^M({1\hspace{-0.243em}\mathrm{l}} + \gamma_*)
          \psi^N\overline\psi^K({1\hspace{-0.243em}\mathrm{l}}
          +\gamma_*)\psi^L R_{KMLN}
          \nonumber
\\
     & & + \varepsilon^{mn} D_i\partial_m A_n^i 
         +\mbox{$\frac {{\mathrm{i}}}{4}$}\overline\psi^M\psi^N \phi^i 
         \partial_N\partial_M D_i  \nonumber
\\
       & & + \mbox{$\frac {1}{2}$}({\mathrm{i}} \overline\psi^N
           \gamma_* \lambda^i  
           -{\mathrm{i}} F^N \phi^i+ \chi_m \gamma^m\psi^N \phi^i) 
           \partial_N D_i ,
\label{2}
\end{eqnarray}
where $e=\det(e_m^a)$, and $h_{mn}$ is the world-sheet metric
built from $e_m^a$. $G_{MN}$, $B_{MN}$ and $D_i$ are arbitrary functions
of the $X^M$, except that $G_{MN}$ and $B_{MN}$ are symmetric and
antisymmetric in $M$ and $N$, respectively,
and $\Gamma_{MNK},H_{MNK}$ and $R_{MNKL}$ are given by
\begin{eqnarray*}
\Gamma_{MNK} & = & \mbox{$\frac 12$} (\partial_M G_{NK} 
                   + \partial_N G_{MK} - \partial_K G_{MN}) \nonumber
\\
H_{MNK} & = & \partial_M B_{NK} + \partial_N B_{KM} + \partial_K B_{MN} 
\nonumber
\\
R_{MNKL} & = & \partial_K\partial_{[M} (G+B)_{N]L} 
               - \partial_L\partial_{[M} (G+B)_{N]K} \nonumber
\end{eqnarray*}
where $\partial_M$ denotes differentiation with respect to $X^M$.
The redefinitions 
$B_{MN}(X)\rightarrow B_{MN}(X)+\partial_{[M}f_{N]}(X)$ and
$D_i(X)\rightarrow D_i(X)+k_i$ modify the Lagrangian only
by total derivatives for any functions $f_N(X)$ and
constants $k_i$. 

Apart from the terms containing fields of the gauge multiplets,
the Lagrangian (\ref{2}) agrees with the one derived in \cite{BRSSS}
when one eliminates the auxiliary fields $F^M$. Hence, the cohomological
analysis shows that the Lagrangian derived in \cite{BRSSS} is in fact unique 
in absence of gauge multiplets (modulo total derivatives, and up to the 
choice of $G_{MN}$ and $B_{MN}$). It should be noted, however, that this 
uniqueness is tied to the gauge transformations (\ref{1}) and may
get lost when one allows that the gauge transformations get
consistently deformed.
E.g., one would expect that the world-sheet diffeomorphisms and 
supersymmetry transformations can be nontrivially deformed such that
the deformed action is invariant under the deformed
transformations if the background has special isometries, by analogy with
the purely bosonic case \cite{BTV2}. Furthermore, there
are certainly actions invariant under nonabelian
gauge transformations of the $A_m^i$ (leading
to nonabelian Born-Infeld actions, among others).
The general deformation problem is currently under study.

\paragraph{Simplified action.}
For further discussion we shall assume in the following that the functions 
$D_i$ coincide with a subset of the fields $X^M$. We denote this subset by
$\{y^i\}$ and the remaining $X$'s by $x^\mu$,
\begin{equation}
\{X^M\}=\{x^\mu,y^i\},\quad D_i=y^i.
\label{3}
\end{equation}
In fact, this assumption is a very mild one because, except at stationary 
points of $D_i(X)$, (\ref{3}) can be achieved by a field redefinition
$X^M\rightarrow\tilde X^M=\tilde X^M(X)$, where this
``coordinate transformation'' is such that each nonconstant $D_i(X)$ becomes 
one of the $\tilde X$'s. Indeed, constant $D_i$ give only constributions
to the Lagrangian which are total derivatives and can thus be neglected, at 
least classically; nonconstant $D_i$ can be assumed to be independent by
a suitable choice of basis for the gauge fields and may thus be taken as 
$\tilde X$'s, at least locally (e.g., if $D_1=D_2$, the Lagrangian depends
only on the combination $A^1_m+A^2_m$ which can be introduced as a new 
gauge field).

It is now easy to see that the Lagrangian (\ref{2}) can actually be simplified
by setting the fields $\psi^i,F^i,\lambda^i,\phi^i$ to zero.
Indeed, owing to (\ref{3}), the classical equations of motion for 
$\lambda^i$ and $\phi^i$ yield $\psi^i=0$ and $F^i=0$. 
The latter equations are algebraic and can be used in the Lagrangian. Then 
the Lagrangian does not contain $\lambda^i$ and $\phi^i$ anymore and the only
remnant of the gauge multiplets are the terms 
$e\varepsilon^{mn}y^i\partial_m A_n^i$. 
This reflects that the gauge multiplets carry no dynamical degrees of 
freedom since the world-sheet is 2-dimensional. Of course, the
transformations (\ref{1}) must be adapted in order to
provide the gauge symmetries of the simplified Lagrangian:
those fields that are eliminated from the action
must also be eliminated from the transformations of the
remaining fields using the equations of motion of the eliminated
fields. This only affects the supersymmetry transformations
of $y^i$ and $A_m^i$. The new supersymmetry transformation
of $y^i$ is then simply zero. This is not in contradiction
with the supersymmetry algebra because the equations of motion
for the $A_m^i$ give $\partial_m y^i=0$
(of course, after eliminating
the fields $\psi^i,F^i,\lambda^i,\phi^i$, the supersymmetry
algebra holds only on-shell). 
The $y^i$ are thus constant on-shell,
their values being integration constants fixed only
by initial conditions. This leads to the interpretation of the $y^i$
as coordinates of ``frozen extra dimensions'' mentioned in the
beginning.

\paragraph{Born-Infeld actions.}
Locally supersymmetric Born-Infeld actions arise from
(\ref{2}) for particular choices of $G_{MN}$ and $B_{MN}$, in
complete analogy to the purely bosonic case \cite{BGS}.
For instance, consider the case with only one gauge field 
($\{A_m^i\}=\{A_m\}$, $\{y^i\}=\{y\}$) and the following
particular choice of $G_{MN}$ and $B_{MN}$,
\begin{eqnarray}
G_{\mu\nu}=\sqrt{1+y^2}\, g_{\mu\nu}(x),& &
B_{\mu\nu}=y\, b_{\mu\nu}(x) \nonumber
\\
G_{yy}=G_{y\mu}=B_{y\mu}=0. &\quad&
\label{5}
\end{eqnarray}
$y$ can be eliminated algebraically.
Eliminating also the world-sheet zweibein $e_m^a$,
the Lagrangian becomes
\begin{eqnarray}
& L=\pm\sqrt{-\det(g_{mn}+{\cal F}_{mn})}+\dots &
\nonumber\\
& g_{mn}=g_{\mu\nu}(x)\partial_m x^\mu\,\partial_n x^\nu& \nonumber
\\
&{\cal F}_{mn}=\partial_m A_n-\partial_n A_m
-b_{\mu\nu}(x)\partial_m x^\mu\,\partial_n x^\nu &
\label{6}
\end{eqnarray}
where we have
assumed $\det(g_{mn})<0$ and $\det(g_{mn}+{\cal F}_{mn})<0$,
and the nonwritten terms involve fermions.
The Born-Infeld Lagrangian 
$L=\pm[+\det(g_{mn}+{\cal F}_{mn})]^{1/2}+\dots$ for 
Euclidean signature ($\det(g_{mn})>0$) corresponds to
$G_{\mu\nu}=(1-y^2)^{1/2}g_{\mu\nu}(x)$.

\paragraph{Global symmetries.}
Our second result concerns the global symmetries of
the action (\ref{2}). These can be obtained from the BRST cohomology
in the space of antifield dependent local functionals with ghost number 
$-1$ \cite{BBH1}. 
We have computed this cohomology completely and present now the resulting
global symmetries for the simplified form of the action 
described above (without the fields $\psi^i,F^i,\lambda^i,\phi^i$
and assuming (\ref{3})). The nontrivial global symmetries 
(a  global symmetry is called trivial when it is equal to a
gauge transformation on-shell) 
are generated by the following transformations,
\begin{eqnarray}
\Delta e_{m}^{~a}&=&0, \quad
\Delta \chi^\alpha_m=0 \nonumber
\\
\Delta X^M &=& {\cal H}^M,\quad
{\cal H}^i=K^i(y),\quad {\cal H}^\mu=V^\mu(X)
\nonumber
\\
\Delta \psi_{\alpha}^{\mu} &=& \psi_{\alpha}^{\nu} \partial_{\nu}V^{\mu}(X)  
\nonumber
\\
\Delta F^{\mu} &=& F^{\nu} \partial_{\nu}V^{\mu}(X)
+\mbox{$\frac 12$}\overline\psi^{\nu}\psi^{\lambda} 
\partial_{\nu}\partial_{\lambda}V^{\mu}(X) \nonumber
\\
\Delta A_m^i &=& b_M^i(X)\partial_mX^M+
a_M^i(X){\varepsilon_m}^n\partial_nX^M
\nonumber\\
&& -\delta_{jk} A_m^j \partial_i K^k(y) 
- \chi_n \gamma_m\gamma^n\gamma_*\psi^{\mu} a_{\mu}^i(X) 
\nonumber\\
&&
+ \mbox{$\frac {{\mathrm{i}}}{2}$} \overline\psi^{\mu}\gamma_m
\{\gamma_*\psi^{\nu} \partial^{}_{[\nu} a_{\mu]}^i(X)
-\psi^{\nu}\partial^{}_{[\nu} b_{\mu]}^i(X)\}  
\label{7}
\end{eqnarray}
where ${\cal H}^M$, $a_M^i$ and $b_M^i$ have to solve
the following generalized Killing vector equations,
\begin{eqnarray}
{\cal L}_{{\cal H}}G_{MN} &=& - 2 \delta_{i(M} a_{N)}^{~i}, \nonumber
\\
{\cal L}_{{\cal H}}B_{MN} &=& - 2\partial_{[M} p_{N]}
                                +2\delta_{i[M} b_{N]}^{~i}
\label{8}
\end{eqnarray}
for some functions $p_M(X)$
(${\cal L}_{{\cal H}}$ is the Lie derivative along 
${\cal H}^M$ and $\delta_{iM}$ 
is the Kronecker symbol). Note that the $p_M$ do not occur in the 
$\Delta$-transformations; however, they do contribute to the corresponding 
Noether currents.

The equations (\ref{8}) are actually the same as the equations
which also determine the symmetries of bosonic string and D-string actions
\cite{BTV,BGS} with the  specification (\ref{3}).
In the absence of gauge fields (no $A_m^i$, $y^i$, $K^i$; 
$\{{\cal H}^M\}\equiv \{V^\mu\}$),
they read
\begin{eqnarray}
{\cal L}_V G_{\mu\nu}=0,\quad
{\cal L}_V B_{\mu\nu}=- 2\partial_{[\mu} p_{\nu]}\ .
\label{9a}
\end{eqnarray}
These equations had been already discussed in \cite{hull}.
The first equation
(\ref{9a}) is just the standard Killing vector equation for
$G_{\mu\nu}$. Hence, the solutions of
equations (\ref{9a}) are those Killing vector fields of $G_{\mu\nu}$
which solve the second equation (\ref{9a}) (for some $p_\mu$).

The situation changes when gauge fields are present. Then 
equations (\ref{8}) read for $M,N=\mu,\nu$:
\begin{equation}
({\cal L}_V+K^i\partial_i) G_{\mu\nu}=0, \
({\cal L}_V+K^i\partial_i) B_{\mu\nu}=-2\partial_{[\mu} p_{\nu]},
\label{9}
\end{equation}
where ${\cal L}_V$ is the Lie derivative along the
vector field $V^M$ given by $V^i=0$, $V^\mu=V^\mu(X)$.
The remaining equations (\ref{8}) just determine the functions
$a_M^i$ and $b_M^i$,
\begin{eqnarray}
& a_{\mu}^i=-{\cal L}_{{\cal H}}G_{\mu i}\ ,\quad
a_i^j=-\mbox{$\frac {1}{2}$} {\cal L}_{{\cal H}}G_{i j}&
\nonumber\\
&b_{\mu}^i=-{\cal L}_{{\cal H}}B_{\mu i}+\partial_i p_\mu\ ,\quad
b_i^j=-\mbox{$\frac {1}{2}$}{\cal L}_{{\cal H}}B_{ij}\ .&
\label{10}
\end{eqnarray}
Here we have used that $p_i$ and 
the parts of $a_i^j$ resp.\ $b_i^j$ which are antisymmetric
resp.\ symmetric in $i,j$ can be set to zero without loss of generality
(the corresponding contributions to $\Delta$ can be removed
by subtracting trivial global symmetries from $\Delta$).

The global symmetries are thus completely determined by equations (\ref{9}). 
Note that these equations reproduce (\ref{9a}) for $K^i=0$, except that now 
$G_{\mu\nu}$ and $B_{\mu\nu}$ depend in general not only on the $x^\mu$ but 
also on the $y^i$. Hence, in general $V^\mu$ and $p_\mu$ also depend 
on the $y^i$. For the discussion of equations (\ref{9}), the 
$y^i$ may be viewed as parameters of $G_{\mu\nu}$ and $B_{\mu\nu}$
rather than as coordinates of extra dimensions. 
Solutions to equations (\ref{9}) with $K^i=0$ can thus be regarded as 
solutions to equations (\ref{9a}) for 
parameter-dependent $G_{\mu\nu}$ and $B_{\mu\nu}$. In contrast, 
solutions to (\ref{9}) with $K^i\neq 0$ have no counterparts among the 
solutions of (\ref{9a}). Such solutions may be called ``dilatational'' 
solutions, because in special cases they are true dilatations, as we will 
see in the example below (further examples can be found in  \cite{BGS}). 

Finally we note that the solutions to equations (\ref{9}) come in infinitely 
big families and that, as a consequence, the corresponding commutator 
algebra of the global symmetries is an infinite dimensional loop-like 
algebra. This has been observed already in \cite{BGS} and is a consequence 
of the fact that the action depends on the $A_m^i$ only via their field 
strengths \cite{BGMS}. All members of a family of solutions arise from one 
solution $V^\mu(X)$, $K^i(y)$, $p_\mu(X)$ by multiplying that solution 
with arbitrary functions of the $y^i$.
As the $y^i$ are constant on-shell, this 
infinite dimensionality of the space of global symmetries has no practical 
importance, i.e., in order to discuss the global symmetries it is sufficient 
to consider just one representative of each family.

\paragraph{Example.}
To illustrate the results presented above, we specify them for a simple class 
of models characterized by Lagrangians containing only one 
$U(1)$ gauge field $A_m$ and the following choices for the background  
\[
G_{y M}=B_{y \mu}=0,
\quad  G_{\mu\nu}=f(y)\eta_{\mu\nu}, 
\quad  B_{\mu\nu}=B_{\mu\nu}(y),
\]
leading to 
\begin{eqnarray*}
L/e &=&  -\mbox{$\frac {1}{2}$} h^{mn} \partial_m x^{\mu} 
         \partial_n x^\nu G_{\mu\nu} 
         + \mbox{$\frac {1}{2}$} \varepsilon^{mn} 
         \partial_m x^{\mu} \partial_n x^\nu B_{\mu\nu}
\\
     & &+ \chi_k \gamma^n \gamma^k \psi^\nu \partial_n x^{\mu} G_{\mu\nu}
      -\mbox{$\frac {1}{4}$}\chi_k\gamma^n \gamma^k C\chi_n
      \overline\psi^{\mu}\psi^\nu G_{\mu\nu} 
\\
     & & +\mbox{$\frac {{\mathrm{i}}}{2}$} \overline\psi^{\mu}
         \gamma^m\partial_m\psi^\nu G_{\mu\nu}  \nonumber
         -\mbox{$\frac {{\mathrm{i}}}{4}$}\overline\psi^\nu
         \gamma^m\gamma_*\psi^{\mu}\partial_m y 
          \,\partial_{y} B_{\mu\nu}
\\
     & & + \mbox{$\frac {1}{2}$}\varepsilon^{mn}(\partial_m A_n 
         - \partial_n A_m)y
\end{eqnarray*}
where the auxiliary fields $F^\mu$ have been eliminated.
As shown in \cite{BGS}, in this case the general solution of equations 
(\ref{8}) is (modulo trivial global symmetries)
\begin{eqnarray*}
K&=&2r(y)
\\
V^{\mu} &=& - r(y) [\ln f(y)]' x^{\mu} 
                  + r^{\mu}(y) + r^{[\mu\lambda]}(y)\eta_{\lambda\nu}x^{\nu}
\\
a_{\mu} &=& - V^{\lambda}{}'f(y)\eta_{\mu\lambda}, \quad a_y = 0
\\
b_{\mu} &=&  (r(y) B'_{\mu\nu})' x^{\nu} 
           + B'_{\mu\nu}V^{\nu}, \quad b_y = 0
\\
p_{\mu} &=& r(y) B'_{\mu\nu} x^{\nu} +  B_{\mu\nu}V^{\nu},
\quad p_y = 0 
\end{eqnarray*}
where a prime denotes differentiation with respect to $y$ and
$r(y)$, $r^{\mu}(y)$ and $r^{[\mu\lambda]}(y)$ are arbitrary
functions of $y$ and correspond to families of dilatations, 
translations and Lorentz-transformations in target space, respectively. For 
three reasons the dilatations are special: (i) as discussed already above,
they have no counterpart among the global symmetries of the ordinary 
superstring on a flat background; 
(ii) they can change the value of the ``extra coordinate'' $y$;
(iii) they can map solutions to the 
classical equations of motion with vanishing field strength 
$\partial_m A_n-\partial_n A_m$ to solutions with non-vanishing field strength,
in contrast to the translations and Lorentz-transformations.
Property (ii) holds because of $\Delta y=2r(y)$ and is intimately
related to property (iii) because the field strength is related to $y$ by 
$f'(y)\approx\varepsilon^{mn}\partial_m A_n+\dots$ where $\approx$ is equality
on-shell. 
We stress that the presence of the dilatations is not
an artefact of our formulation. 
Rather, they are of course present even after elimination of $y$.
Finally we note that properties (ii) and (iii) extend to more 
complicated backgrounds for which
solutions to (\ref{9}) with $K^i\neq 0$ exist.

\paragraph{Acknowledgements.} FB thanks the Erwin-Schr\"odinger-Institute for 
hospitality and financial support during the time when this work was completed,
and was supported by a DFG habilitation grant at earlier stages of the work.
AK was supported by \"ONB under project grant number 7731 and by ``Fonds zur 
F\"orderung der Wissenschaftlichen Forschung'' under project grant number 
P13125-TPH.

\end{document}